\documentclass{PoS}

\title {Exact Chiral Fermions and Finite Density on Lattice}

\ShortTitle{Chiral Fermions \& Finite Density}

\author{Debasish Banerjee \\
        Department of Theoretical Physics, Tata Institute of Fundamental
        Research,\\ Homi Bhabha Road, Mumbai 400005, India \\
        E-mail:\email{debasish@theory.tifr.res.in}}

\author{\speaker{Rajiv V. Gavai} \thanks{Supported by the Indian
Lattice Gauge Theory Initiative project of TIFR, Mumbai, India.}\\
        Department of Theoretical Physics, Tata Institute of Fundamental
        Research,\\ Homi Bhabha Road, Mumbai 400005, India \\
        E-mail:\email{gavai@tifr.res.in}}

\author{Sayantan Sharma \thanks{Supported by the Shyamaprasad Mukherjee Fellowship
of CSIR, India.}\\
        Department of Theoretical Physics, Tata Institute of Fundamental
        Research,\\ Homi Bhabha Road, Mumbai 400005, India \\
        E-mail:\email{ssharma@theory.tifr.res.in}}

\abstract{ Any $\mu^2$-divergence is shown analytically to be absent for a class
of actions for Overlap and Domain Wall Fermions with nonzero chemical
potential.  All such actions are, however, shown to violate the chiral
invariance. While the parameter M of these actions can be shown to be
irrelevant in the continuum limit, as expected, it is shown numerically that
the continuum limit can be reached with relatively coarser lattices for 1.5 
$\leq M \leq $1.6.  }

\FullConference{The XXVI International Symposium on Lattice Field Theory\\
		 July 14-19 2008\\
		 Williamsburg, Virginia, USA}

\begin{document}

\section{Introduction.}

Quantum Chromo Dynamics (QCD) with massless quarks of $N_f$ flavours has
$SU(N_f) \times SU(N_f)$ chiral symmetry which is dynamically broken at low
temperatures by the vacuum.   Since the $u$ and $d$ quarks have almost
degenerate mass which in turn is much smaller than the scale of QCD, and the
$s$ quark is only moderately heavy, approximate chiral symmetry is a reasonable
assumption for our world, i.e., QCD with 2 + 1 flavours of dynamical quarks.
Rich phenomenological studies have demonstrated the utility, and consistency of
such an assumption.  Pions and Kaons are thus widely regarded as the Goldstone
bosons resulting from the dynamical symmetry breaking by vacuum.  It is thus a
natural consequence that the finite temperature transition in our world is also
widely accepted to be governed by chiral symmetry.  It is therefore desirable
to have the chiral invariance in any formulation which aims at studying the
thermodynamics of QCD.  Let us cite below two important physics aspects where
it is even necessary to have exact chiral invariance.
 
Lattice QCD has clearly been the method of choice for reliable non-perturbative
studies in general, and finite temperature/density investigations in
particular.   As is well known, the Fermion doubling problem makes it difficult
to demand the same chiral invariance as in continuum QCD.   The original
solution to this problem, namely, the Wilson Fermions, breaks all chiral
symmetries.  Since the staggered Fermions do have an exact chiral symmetry on
the lattice, albeit at the cost of breaking of flavour and spin symmetries,
they have dominated the area of nonzero temperatures and densities.  

As presented \cite{lat06} in Lattice 2006 by one of us, the hadronic screening
lengths illustrate their deficiency in the pionic screening length whereas the
Overlap Fermions, with exact chiral, flavour and spin symmetry for any
arbitrary lattice spacing, appear to do better.  Advocated as means to explore
the large scale composition of QGP, various numerical investigations with
staggered quarks found that except the pion (and the sigma) screening length,
all others could be understood as multiples of the appropriate number of free
quarks (or antiquarks).  The pionic correlator on the other hand showed
nontrivial structure, and the pionic screening length approached the ideal gas
value only in the continuum limit.  The simulations with overlap quarks yielded
a pleasant surprise on both counts and the corresponding pion screening length
was close to ideal gas value for even small temporal lattices.

Another fundamental aspect of QCD is the existence and location of the critical
point in the $T$-$\mu_B$ plane, where $\mu_B$ is the chemical potential for
baryon number. Based on symmetries and a variety models, the QCD phase diagram
is expected to have a critical point for two light and one moderately heavy
quark.  Again chiral symmetry plays a crucial role in this : the transition at
$\mu_B =   0$ should be second order for two massless quarks, which turns into
a cross over for light quarks.  A line of first order phase transitions at
finite density ought to terminate in a critical point.

\section{Ginsparg-Wilson Relation and $\mu \ne 0$.}

Exact chiral invariance for a lattice Fermion operator $D$ is assured if it
satisfies the Ginsparg-Wilson relation \cite{gw} : $ \{ \gamma_5, D \} = a D
\gamma_5 D$.  In particular, the chiral transformations \cite{lus} $ \delta
\psi = \alpha  \gamma_5(1 - \frac{a}{2}D ) \psi$ and $ \delta \bar \psi =
\alpha \bar \psi (1 - \frac{a}{2}D )\gamma_5 $, leave the action $S =
\sum_{x,y} \bar \psi(x) D_{x,y} \psi(y)$ invariant: 
\begin{equation}
\delta S = \alpha~~ \sum_{x,y} \bar \psi_x \big[ \gamma_5 D +
D \gamma_5 - \frac{a}{2}  D \gamma_5 D
-\frac{a}{2} D \gamma_5 D \big]_{xy} \psi_y = 0
\end{equation}
The Overlap Fermions, and the Domain Wall Fermions in the limit of large
fifth dimension satisfy this relation. Thus these Fermions with exact
chiral invariance on the lattice are ideal for studies of the QCD phase
diagram.  One needs to introduce chemical potential in their known actions
to do so.  For the staggered quarks, this was done by first finding an
expression for the conserved number, $N$, on the lattice, and then adding
$\mu N$ term to the action.  It turned out \cite{bg} that this lead to
$\mu^2$-dependent divergences in the continuum limit even for the free
case, which were removed by further modification of the action. Since the
non-locality of the Overlap or Domain Wall Fermions makes the construction
of a conserved charge difficult, it was proposed \cite{bw} that the Wilson
Dirac operator $D_W$ in the actions for these Fermions be modified using
the same prescription as in the staggered case.  This amounted to
multiplying the link variable in the positive (negative) time direction by
$K(a\mu) = \exp(a\mu)$ ($L(a\mu)  = \exp(- a\mu)$).  As the  $ \gamma_5 D_W
(a\mu) $ is no longer Hermitian, the $D_{ov} = 1+ \gamma_5 sgn(\gamma_5
D_W)$ definition of the overlap operator necessitated an extension of the
sign function:  For complex $ \lambda = (x + iy)$ eigenvalue, $
sgn(\lambda) = sgn(x)$. Using this operator, it was shown \cite{GatLip}
showed numerically that no $\mu^2$-divergences exist in the free case
($U=1$).  We demonstrated \cite{bgs} the absence of the divergence in the
free case analytically for all $K,L$ such that $K(a\mu) \cdot L(a\mu) = 1$. 

We also claim \cite{bgs} though that the chiral invariance is lost for
nonzero $\mu$.  This is easy to see by varying the quark fields again by
the same infinitesimal chiral transformation as above :
\begin{equation}
\delta S (a\mu)  = \alpha \sum_{x,y} \bar \psi_x \big[ \gamma_5 D(a\mu) +
D(a\mu) \gamma_5 - \frac{a}{2}  D(0) \gamma_5 D
(a\mu) -\frac{a}{2} D(a\mu) \gamma_5 D(0) \big]_{xy}
\psi_y ~,~
\label{dsmu}
\end{equation}
\noindent whereas the extended sign function definition of the Dirac
operator merely ensures 
\begin{equation}
\gamma_5 D(a\mu) + D(a \mu) \gamma_5 -a~ D(a\mu) \gamma_5
D(a\mu) = 0 ~.~
\end{equation}
This is clearly  not sufficient to make $\delta S = 0$.
This is true for both Overlap and Domain Wall Fermions and for any $K$,$L$. 

A direct consequence is that the much desired  exact chiral symmetry on
lattice is lost for any $\mu \ne 0$, real or imaginary. Thus a
$\mu$-dependent mass will be acquired by even massless quarks in the
interacting theory.  The behaviour of chiral condensate as a function of
$\mu$ will therefore be necessarily smoothened, wiping out any chiral
transition that may be present.   Of course, depending on how strong the
transition is, it will begin to show up for small enough $a$ or large enough
$N_T$.   How large a computational effort that may mean is a priori not
clear.   Recall that in the Taylor expansion method \cite{gg} to
incorporate the effects of nonzero $\mu$, all the coefficients are
evaluated at $\mu=0$.  These will not suffer from any such lack of chiral
invariance but the series in $\mu$ will be smooth, and will always exhibit
convergence for any $\mu$.

One may be tempted to modify the chiral transformation itself for nonzero
$\mu$ by demanding  $ \delta \psi = \alpha  \gamma_5(1 - \frac{a}{2}D (a
\mu )) \psi$ and $ \delta \bar \psi = \alpha \bar \psi (1 - \frac{a}{2}D (a
\mu))\gamma_5 $.   Clearly, $\delta S(a\mu) = 0$ in that case.  This is,
however, not permissible since $ \gamma_5 D(a \mu)$, i.e, the generator of
the transformation is not Hermitian.  Moreover, a symmetry transformations 
should not depend on the ``external'' tunable parameter $\mu$.  Recall that
the chemical potential is introduced for charges $N_i$ with $[H, N_i] =0$.
At least, the symmetry should therefore not change as $\mu$ does.  The most
damaging practical consequence of such a modification of the chiral
transformation is that the restoration of chiral symmetry due to an
increase in $\mu$ cannot be addressed at all.  For $T \ne 0$ with $\mu=0$, 
the symmetry group remains the {\em same} at each $T$, allowing a change in the
chiral order parameter  $\langle \bar \psi \psi \rangle (am=0, T)$ to be
interpreted as a change in the vacuum, i.e, restoration of the dynamically
broken symmetry.  With $\mu$-dependent chiral transformations, the symmetry 
groups are {\em different} at each $\mu$, with no obvious relation between
the respective chiral condensates.   It thus appears much more reasonable
{\it not} to alter the transformations but to look for a better way to
include the chemical potential in the Overlap Dirac operator.

\section{Free Overlap and Domain Wall Fermions: Analytical Results.}

\begin{figure}[htb]
\vskip -0.3 cm
\includegraphics[scale=0.27]{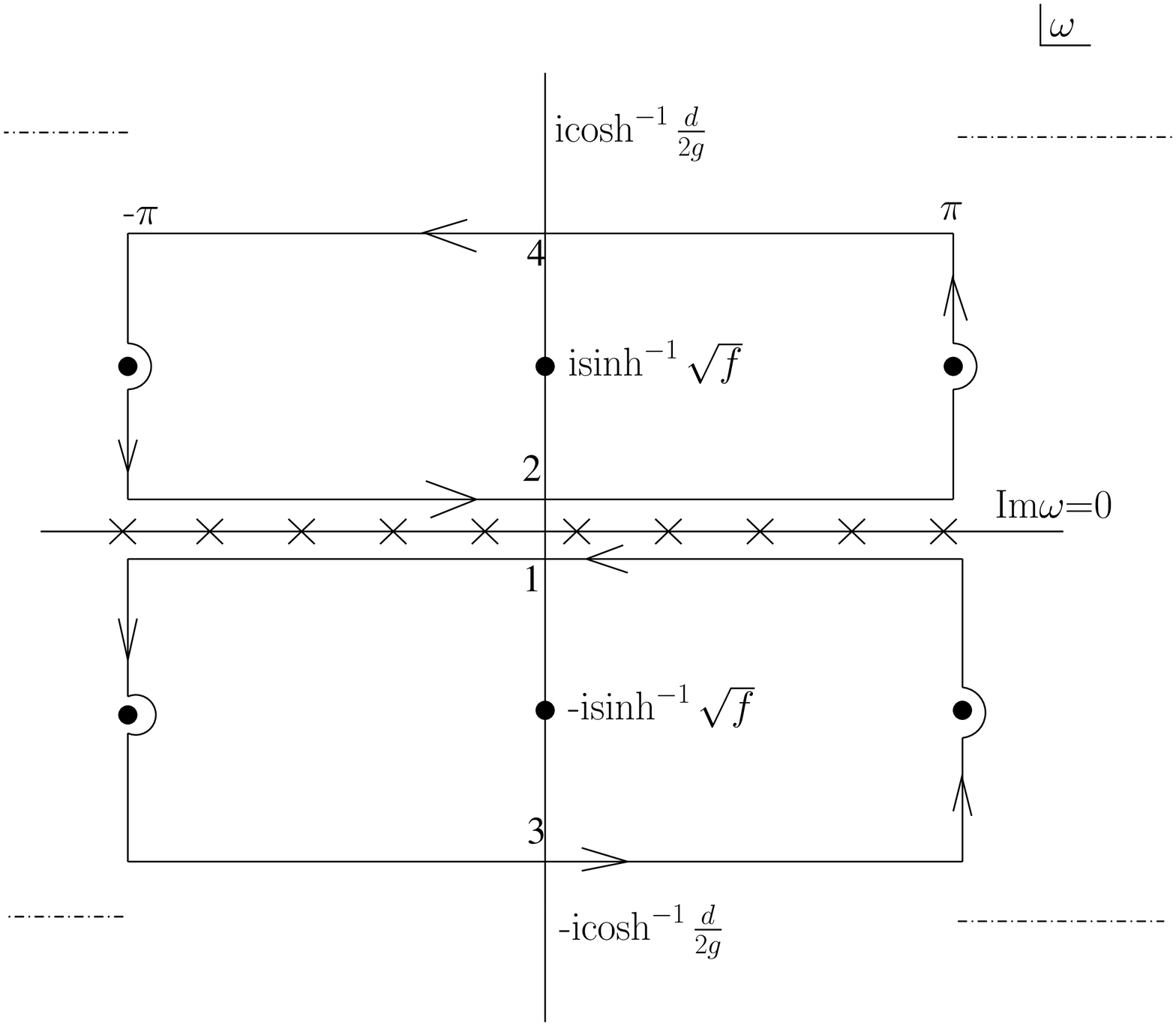}
\includegraphics[scale=0.30]{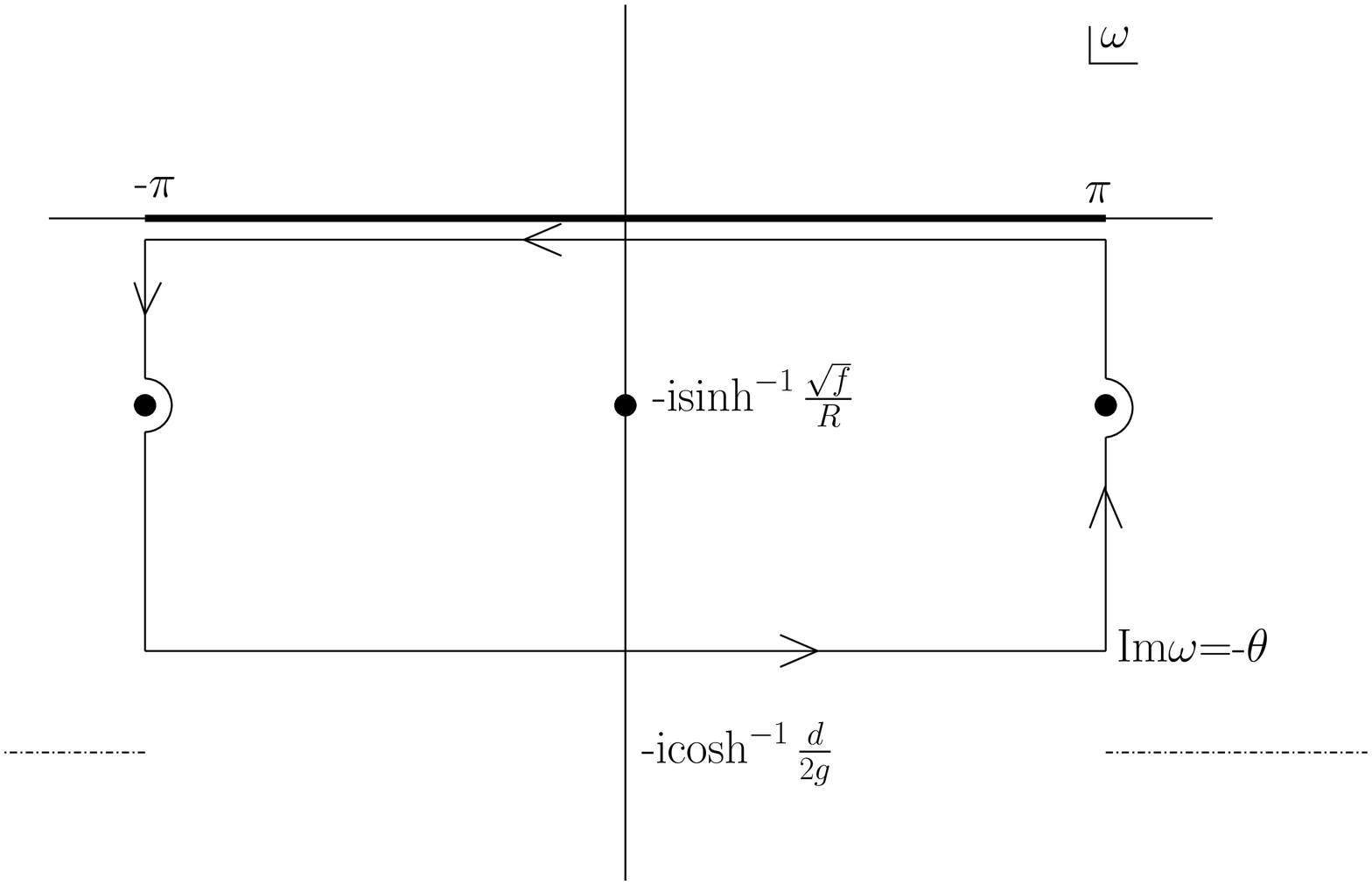}
\vskip -0.3 cm
\caption{Contours in the complex $\omega$-plane for $\mu =0$ (left) and
$T=0$(right) cases.}
\label{figcon}
\vskip -0.3 cm
\end{figure} 

We investigated thermodynamics of free overlap and domain wall Fermions
both analytically and numerically to demonstrate that i) the negative mass 
parameter $ 0 < M < 2 $ is irrelevant in the continuum limit, and ii) the
absence of $\mu^2$-divergences for general $K$ and $L$.  Using our
numerical results, we also obtained an optimal range for $M$ in order to
obtain small deviations from the continuum limit on coarse lattices.

The energy density, pressure, quark number susceptibility etc. can be obtained 
from $ \ln {\cal Z} = \ln~{\rm det}~D_{ov}$ by taking $T$, $V$, and $\mu$ or 
equivalently $a_4$ and $a$ and $a\mu$, partial derivatives.  Here $V =
N_s^3 a^3$ and $T = 1/(N_T a_4)$. Since the Dirac operator is diagonal in 
momentum space, its eigenvalues can be used to compute ${\cal Z}$:
\begin{eqnarray}
\nonumber
\lambda_{\pm}&=&1-[sgn\left(\sqrt{h^2+h_5^2}\right) h_5 \pm i \sqrt{h^2}]/\sqrt{h^2+h_5^2}~,~ 
{\rm ~with~~} \\ \nonumber
h^2 = \sum_{i=1}^4 h_i^2,~ h_j &=& - \sin ap_j,~ j =1, 2 {\rm~and~} 3,~ h_4 =
-a~\sin(a_4p_4)/a_4 \\ 
{\rm~and~} h_5 &=& M - \sum_j^3 [1 - \cos(ap_j)] -a [1-\cos(a_4p_4)]/a_4~.~
\label{hdef}
\end{eqnarray}

It is straightforward to show that $\epsilon = 3P$ for all $a$ and $a_4$.
Note that the free energy on finite volume is in general not equal to the
pressure $|P|$, and thus will have {\cal O}(1/V) corrections spoiling the
equality above.   Hiding the spatial momentum $p_j$-dependence in terms of 
known functions \cite{bgs} $g$, $d$ and $f$, the energy density on an $N_s^3 
\times N_T$ lattice is 

\begin{eqnarray}
\epsilon a^4 = \frac{2}{N_s^3 N_T}\sum_{p_j,n} F(1, \omega_n) & =&
\frac{2}{N_s^3 N_T}
\sum_{p_j,n} \left[(g+ \cos\omega_n)+ \sqrt{d+2 g \cos\omega_n}\right] \\ \nonumber
&\times& \left[ \frac{(1-\cos\omega_n) }{d+2g \cos\omega_n} + 
\frac{\sin^2\omega_n(g+\cos\omega_n)}{(d+2g \cos\omega_n)(f+\sin^2\omega_n)} \right]~.~ 
\label{endn}
\end{eqnarray}
where $\omega_n (=a_4p_4) $ are the Matsubara frequencies.  As in the case
of the continuum, sum over $\omega_n$ can be carried out using the contour
technique in the complex $\omega$-plane.  The left panel in the Figure \ref{figcon} 
displays the contour chosen for the $\mu=0$ case. The crosses denote the
Matsubara frequencies, the circles denote the physical poles in eq.
(\ref{endn}) corresponding to the zero of $ (f+\sin^2\omega)$ and the
dashed lines correspond to the cuts.  Evaluating the integrals, one obtains
$ \epsilon a^4 =4N_s^{-3} \sum_{p_j}\left [\sqrt{f/1+f}\right] [\exp(N_T\sinh^{-1}
\sqrt{f})+1]^{-1} + \epsilon_3+\epsilon_4$~,~ where $f =  \sum_j \sin^2
(ap_j)$, and the contributions from the contours at the top and bottom are
denoted by $\epsilon_{3,4}$ respectively.  In the continuum limit, the
above expression reduces to the ideal gas energy density $\epsilon_{SB}$
for all $M$, with the cuts moving away to infinity faster than the contours
at the to and bottom.

For the case of $T=0$ but $\mu \ne 0$, the contour is displayed in the
right panel of Figure \ref{figcon}. Essentially the same treatment goes
through as above, if one defines  $K(\mu)+L(\mu) = 2 R \cosh \theta$ and 
$K(\mu)-L(\mu) =  2 R \sinh \theta$ and substitutes in eq. (\ref{endn}) 
$R \sin ( \omega_n - i \theta)$ for $\sin \omega_n$ and similarly for
$\cos \omega_n$.  After the contour integral one obtains 
\begin{equation}
\epsilon a^4 = \frac{1}{\pi N_s^3}\sum_{p_j}\left[2\pi {\rm Res}~ F(R, \omega) 
\Theta\left(K(a\mu)-L(a\mu)-2\sqrt{f}\right) \right. 
\left . +
\int_{-\pi}^{\pi}F(R,\omega)d\omega-\int_{-\pi}^{\pi}F(1,\omega)d\omega\right].
\end{equation}

From the above expression, one notices that $R = K(a\mu) \cdot L(a\mu) = 1$
ensures cancellation of the last two terms.  For $R \ne 1$, expanding $R$
in powers of $\mu$, on the other hand, the $\mu^2$-divergence is explicitly
seen in the continuum limit.  Note the higher order terms in that case
violate the Fermi surface condition in the $\Theta$-function and contribute
for all $\mu$ on the lattice; they vanish in the continuum limit.  From the
$\Theta$-function in the expression above, it is also clear that $K$ and
$L$ should be such that $K(a\mu) - L(a\mu) = 2 a~\mu + {\cal O}(a^3)$ in
order that the usual Fermi surface condition is recovered.  Furthermore,
$K(0) =1 = L(0)$ is necessary for consistency with the zero density Dirac
operator.   We have also shown that the above derivations go through for
the Domain Wall Fermions \cite{sg} and for the general case of $T\ne0$ and
$\mu \ne 0$ for both the Overlap \cite{bgs} and Domain wall Fermions.

\section{Free Overlap and Domain Wall Fermions: Numerical Results.}

\begin{figure}[htb] 
\begin{center}
\includegraphics[scale=0.8]{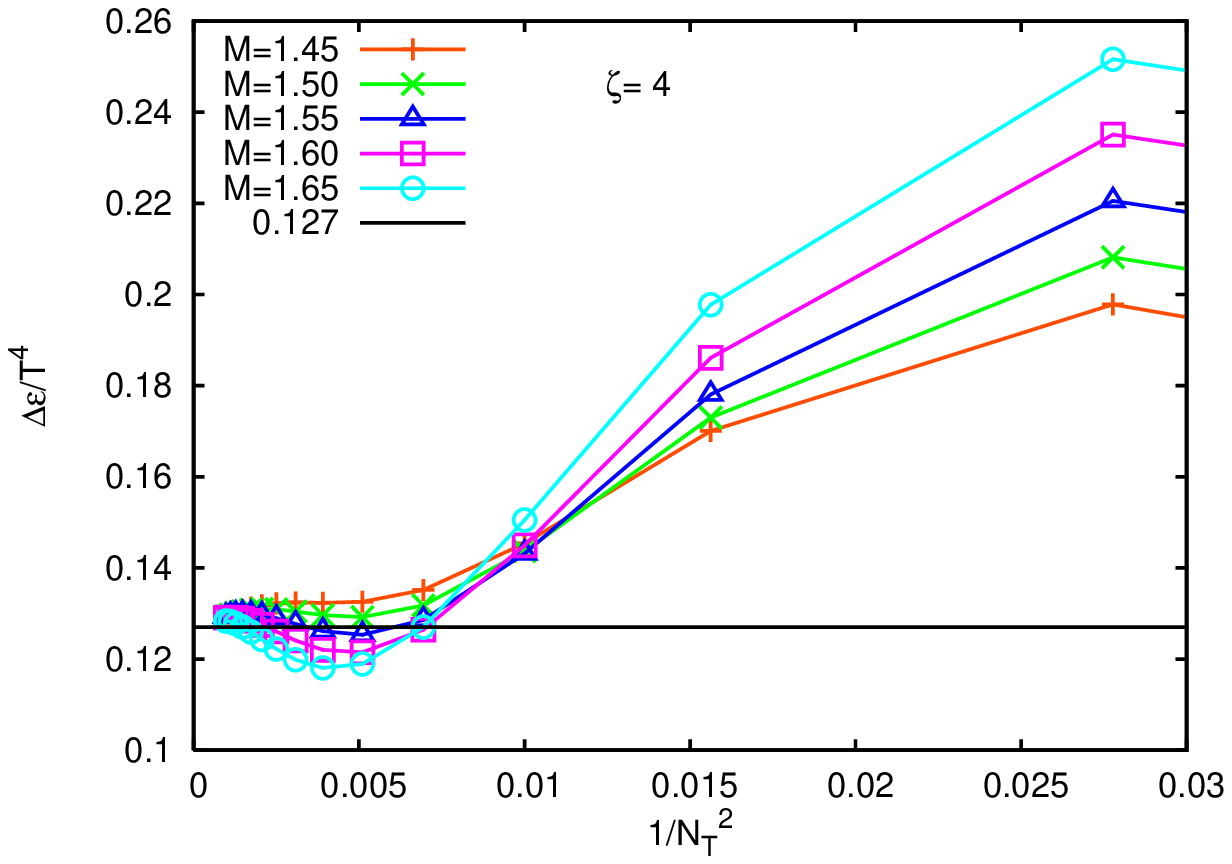}
\includegraphics[scale=0.8]{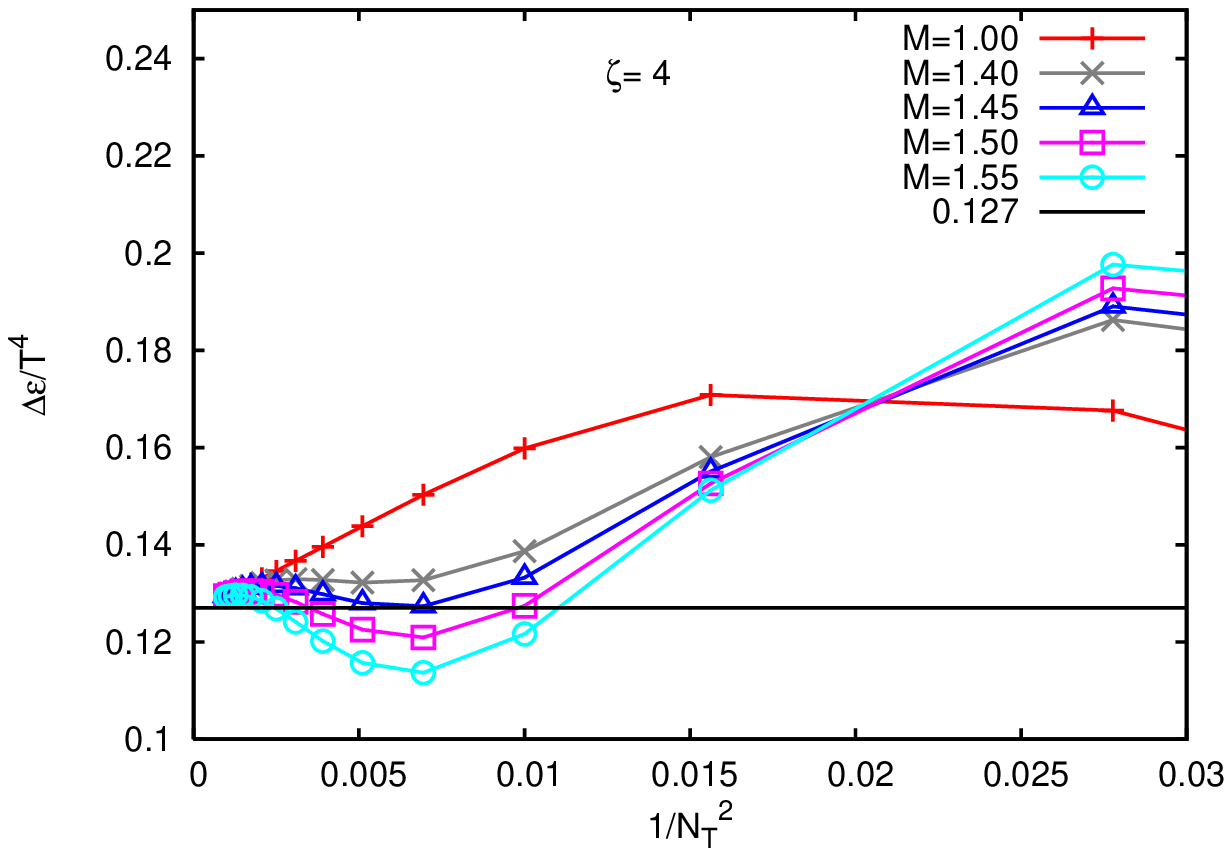}
\end{center}
\caption{ $\mu$-dependent contribution to the energy density for Overlap
(top) and Domain Wall (bottom) Fermions for $\mu/T=0.5$.}
\label{figdel}
\end{figure} 

Numerical evaluation of the physical observables is a simple evaluation of
sums over all allowed momenta $a p_j = (2 \pi /N_s) n_j$ with $n_j=0,~N_s-1$ and
$ a_4 p_4 = (2  n_4 + 1) \pi/N_T $ with $n_4=0,~N_T-1$.  Fixing  a large $
\zeta = LT = N_s/N_T$ for thermodynamic limit, we let $N_T \to \infty$ to
obtain the results in the continuum limit. Figure \ref{figdel} shows 
$\Delta \epsilon (\mu,T)/T^4 = [\epsilon(\mu,T) - \epsilon(0,T)]/T^4$ for $\mu/T
=0.5$ for the Overlap (top panel) and the Domain Wall (bottom) Fermions
for $\zeta=4$.  The  $\zeta = 5$ results have been checked to be
indistinguishable from these. Both sets of results show small, $\sim 2$-3
\%, deviations from the continuum result (= 0.127 for $\mu/T=0.5$) for $N_T 
\ge 12 $ for a range of $M$, being  $1.50 \leq M\leq1.60$ for Overlap
Fermions and $1.40 \leq M\leq1.50$ for the Domain Wall Fermions.  Note that
the canonical $M=1$ choice yields a smoother $1/N_T$ dependence but larger
deviations.  

\begin{figure}[htb]
\begin{center}
\includegraphics[scale=0.8]{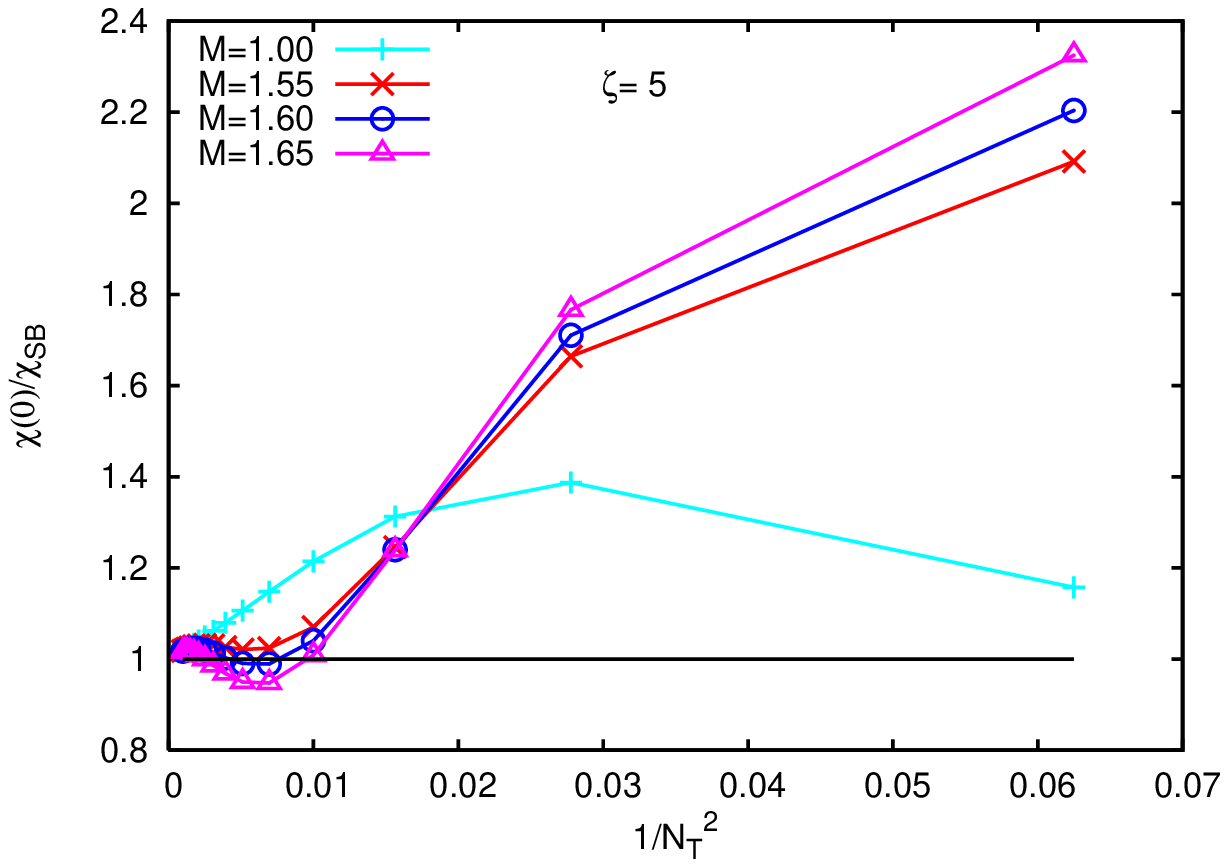}
\includegraphics[scale=0.8]{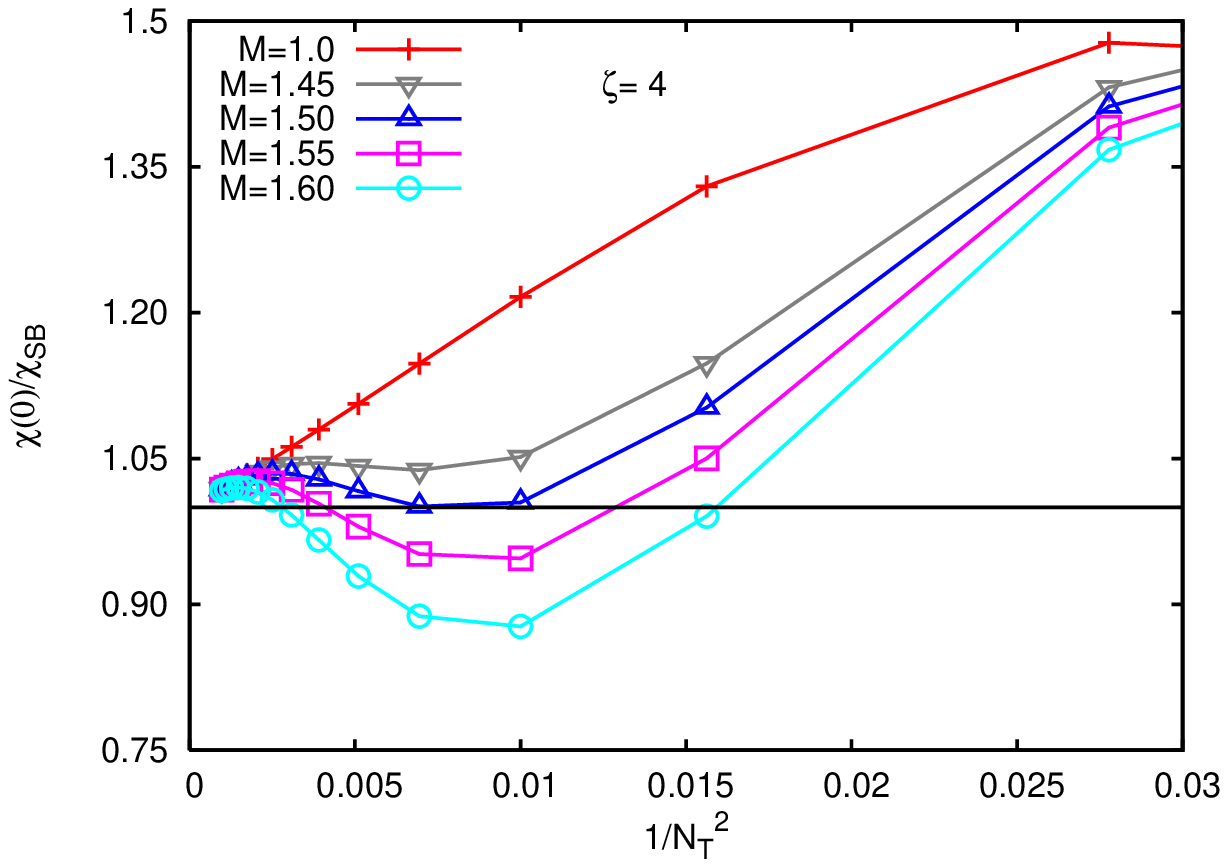}
\end{center}
\caption{ Quark number susceptibility as a function of $1/N_T^2$ for Overlap
(top) and Domain Wall (bottom) Fermions for $\mu=0$.}
\label{figsus}
\end{figure}

Quark number susceptibility, defined as 
\begin{equation}
\chi=\frac{T}{V}\left(\frac{\partial^2 ln \det D}{\partial
\mu^2  }\right)_{T,V}~,~
\end{equation}
can be worked out for the Overlap Fermions to be
\begin{equation}
\chi = \frac{2i}{N_T N_s^3 a^2}\sum_{p_j,p_4}\left[\frac{-\left(h^2 h_4+h_4 h_5 \cos(ap_4-i  a \mu )\right)u}{s^4(s-h_5)^2}  + \frac{ v}{s^2(s-h_5)}\right]~.~
\end{equation}
The u and v in the expression above are
\begin{eqnarray}
 u &=& 2(s-h_5)(h_4\frac{\partial h_4}{\partial a\mu }+h_5\frac{\partial h_5}{\partial
a \mu })  + s^2(\frac{\partial s}{\partial a \mu}-\frac{\partial h_5}{\partial
a \mu})~,~  {~~\rm and~~} \\ \nonumber
 v&=&\frac{\partial h_4}{\partial a \mu }(2h_4^2+h^2+h_5\cos (ap_4-i a \mu ))
 +  h_4\frac{\partial h_5}{\partial a \mu}\cos(ap_4-i  a \mu)+i h_4
h_5\sin (ap_4-i a \mu )~,~
\end{eqnarray}
with $ s^2=h^2+h_5^2$ and the $h_i$ as defined in eq.(\ref{hdef}) but with
$a_4p_4 \to (ap_4 - i a \mu)$. Similarly one can write down an expression
for it for the Domain Wall Fermions.  Figure \ref{figsus} displays the
results for the Overlap Fermions (top panel) and the Domain Wall Fermions
(bottom panel) as a function of $1/N_T^2$. The behaviour is each case is
similar to the corresponding energy density case in Figure \ref{figdel}.
In particular, the same range of $M$ seems to be optimal for both
$\Delta \epsilon$ and $\chi (\mu=0)$.

\section{Summary.}

Exact chiral symmetry for the quark fields on the lattice, without any
violation of other symmetries, such as the flavour or spin, is important
for several investigations of QCD thermodynamics.  Indeed, the existence of
the critical point in the $T$-$\mu_B$ phase diagram of QCD, and the
non-perturbative determination of its location, depend on it.  The
currently popular choice of staggered Fermions may not be fully adequate
for such studies.  The Overlap and Domain wall Fermions are ideally suited
but unfortunately lose their chiral invariance on introduction of chemical
potential in the Bloch-Wettig method and its generalizations, as we showed
above.  We also proved analytically that no $\mu^2$-divergence exists in
the continuum limit for both the Overlap and Domain Wall Fermions for the
Bloch-Wetting method and an associated  general class of functions $K(\mu)$
and $L(\mu)$ with $K(\mu) \cdot L(\mu) = 1$.  Our numerical results
corroborate these findings, and lead us to an optimal range for the
irrelevant parameter $M$. For the choice of $ 1.5 \le M \le 1.6$ ($ 1.4 \le
M \le 1.5$), both the quark number susceptibility at $\mu=0$ and the
$\mu$-dependent part of the energy density exhibited the smallest
deviations from the ideal gas limit for $N_T \ge 12$ for Overlap (Domain
Wall) Fermions.

\section{Acknowledgements.}
One of us (R.V.G.) wishes to thank the members of the local organizing
committee for their kind hospitality which enabled his participation in the symposium.

\end{document}